\documentclass[sigconf,natbib]{acmart}
\AtBeginDocument{%
  \providecommand\BibTeX{{%
    \normalfont B\kern-0.5em{\scshape i\kern-0.25em b}\kern-0.8em\TeX}}}

\usepackage{graphicx}
\usepackage{amsmath}
\usepackage{caption}
\usepackage{subcaption}
\usepackage{hyperref}
\hypersetup{
colorlinks=true,
linkcolor=black}
\usepackage{xcolor}
\usepackage{multirow}
\usepackage{multicol}
\usepackage{balance}

\copyrightyear{2023}
\acmYear{2023}
\setcopyright{acmlicensed}\acmConference[SIGIR '23]{Proceedings of the 46th International ACM SIGIR Conference on Research and Development in Information Retrieval}{July 23--27, 2023}{Taipei, Taiwan}
\acmBooktitle{Proceedings of the 46th International ACM SIGIR Conference on Research and Development in Information Retrieval (SIGIR '23), July 23--27, 2023, Taipei, Taiwan}
\acmPrice{15.00}
\acmDOI{10.1145/3539618.3591963}
\acmISBN{978-1-4503-9408-6/23/07}

\begin{document}

\title{Click-Conversion Multi-Task Model with Position Bias Mitigation for Sponsored Search in eCommerce}

\settopmatter{authorsperrow=4}

\author{Yibo Wang}
\email{ywang633@uic.edu}
\orcid{0000-0001-8872-1811}
\affiliation{
  \institution{\mbox{\!\!\!\!\!University of Illinois Chicago}}
  \city{Chicago}
  \state{IL}
  \country{USA}
  \postcode{60607}
}
\author{Yanbing Xue}
\email{yanbing.xue@walmart.com}
\affiliation{%
  \institution{Walmart eCommerce}
  \city{Sunnyvale}
  \state{CA}
  \country{USA}
}
\author{Bo Liu}
\email{bo.liu1@walmart.com}
\affiliation{%
  \institution{Walmart eCommerce}
  \city{Sunnyvale}
  \state{CA}
  \country{USA}
}
\author{Musen Wen}
\email{musen.wen@walmart.com}
\affiliation{%
  \institution{Walmart eCommerce}
  \city{Sunnyvale}
  \state{CA}
  \country{USA}
}
\author{Wenting Zhao}
\email{wzhao41@uic.edu}
\affiliation{%
  \institution{\mbox{\!\!\!\!\!\!University of Illinois Chicago}}
  \city{Chicago}
  \state{IL}
  \country{USA}
  \postcode{60607}
}
\author{Stephen Guo}
\email{sguo@indeed.com}
\affiliation{%
  \institution{Indeed}
  \city{Sunnyvale}
  \state{CA}
  \country{USA}
}
\author{Philip S. Yu}
\email{psyu@uic.edu}
\affiliation{%
  \institution{\mbox{\!\!\!\!\!\!University of Illinois Chicago}}
  \city{Chicago}
  \state{IL}
  \country{USA}
  \postcode{60607}
}

\renewcommand{\shortauthors}{Yibo Wang et al.}


\begin{abstract}
Position bias, the phenomenon whereby users tend to focus on higher-ranked items of the search result list regardless of the actual relevance to queries, is prevailing in many ranking systems. Position bias in training data biases the ranking model, leading to increasingly unfair item rankings, click-through-rate (CTR), and conversion rate (CVR) predictions. 
To jointly mitigate position bias in both item CTR and CVR prediction, we propose two position-bias-free CTR and CVR prediction models: Position-Aware Click-Conversion (PACC) and PACC via Position Embedding (PACC-PE). PACC is built upon probability decomposition and models position information as a probability. PACC-PE utilizes neural networks to model product-specific position information as embedding. Experiments on the E-commerce sponsored product search dataset show that our proposed models have better ranking effectiveness and can greatly alleviate position bias in both CTR and CVR prediction.
\end{abstract}
\vspace{-1cm}


\vspace{-1cm}
\keywords{Multi-task Learning; Sponsored Product Search; Position Bias}



\maketitle

\vspace{-0.1cm}
\section{Introduction}
The phenomenon of position bias is commonly observed in many ranking and information retrieval systems, including digital advertising and recommender systems. 
Position bias is the tendency of users to pay greater attention to higher-ranked items, regardless of their actual relevance to the query.
For example, the eye-tracking study~\cite{eyetracking1} and~\cite{eyetracking2} show that in web search, the highest-ranked items receive the most attention; the study of digital library recommendation system~\cite{library} discovers that items shown at the top positions are more often clicked regardless of their actual relevance.

A common practice from machine-learned ranking (MLR) models is to use implicit user feedback, such as click/no-click as training data labels. However, due to inherited position bias, these models tend to exhibit lower predicted click-through rates (CTR) for 
lower-ranked items.
This bias in prediction can then 
affect subsequent training data collection,
leading to a persistent and cyclical effect.

The accumulative position bias of the training data will skew the MLR model, leading to increasingly unfair item rank prediction~\cite{craswell2008experimental}.
To address this position bias problem, various approaches have been proposed, including factorization models~\cite{chen2012position}, inverse propensity scores~\cite{ubiased_ltr, metrics,guo2020debiasing}, deep neural network-based CTR prediction models~\cite{ling2017model,guo2019pal,zhuang2021cross}, and more. 
Pioneering approaches such as~\cite{chen2012position} proposed a factorization model that decouples CTR into position-normalized CTR and position bias, which are then estimated by an Expectation-Maximization (EM) algorithm framework.
To remove the position bias for learning to rank models, a propensity-weighted empirical risk minimization framework is proposed in~\cite{ubiased_ltr}. In~\cite{hu2019unbiased}, an unbiased LambdaMART model is proposed, which jointly estimates the biases at click positions and unclick positions and learns an unbiased ranker. Several literatures dedicate to the propensity estimation, such as~\cite{ai2018unbiased,vardasbi2020cascade,agarwal2019estimating}. 
Recently, increasing research studies correcting position bias in the deep-ranking models. 
In~\cite{jin2020deep}, the authors propose to combine recurrent neural network and survival analysis techniques to model unbiased user behaviors. 
~\cite{chen2020context} designs a neural-based Context-Aware Click Model with an examination predictor able to automatically learn the position bias during training. 

The existing methods primarily
consider single-task objectives like CTR prediction.
However, in E-commerce, typical ranking models can have multiple objectives, such as maximizing both CTR and CVR.
Deep Multifaceted Transformers (DMT)~\cite{dmt} learns both CTR and CVR predictors by modeling multiple user behaviors simultaneously with 
bias mitigation and treats CTR and CVR prediction as two parallel tasks. However, item impression, click, and conversion processes have sequential dependencies and are all affected by position bias. Such dependencies are not leveraged in DMT.

In this paper, two models, 
Position Aware Click-Conversion (PACC) and PACC with Position Embedding (PACC-PE) are proposed to model the sequential relationship and jointly mitigate position bias in both CTR prediction and CVR prediction. 
PACC is based on the following two assumptions: 
(1). whether an item will be seen is only related to its position; (2). after an item is seen, whether it will be clicked/purchased is independent of its position.
PACC is built upon the probability decomposition presented in \S\ref{ssect:theory}.
PACC-PE is a variant of PACC, which adopts a neural network to model the product-specific position information as embedding. Compared to PACC, learning the product-specific position embedding with a neural network enables PACC-PE to achieve richer information and superior performance than PACC.

Specifically, our work has the following contributions: 
\begin{itemize}
    \item We propose to jointly learn position-bias-free CTR and CVR prediction models in a multi-task learning framework. By mitigating position bias, the proposed models achieve comparable performance as state-of-the-art models on CTR prediction and significant performance improvement on CVR prediction regarding weighted Mean-Reciprocal-Rank (MRR)~\cite{metrics}, MRR, position-wise AUC (PAUC)~\cite{huang2021metrics2}, and AUC.
    \item We conduct experiments on real-world E-commerce sponsored product searches. Our proposed models achieve better ranking effectiveness and greatly mitigate position bias.
\end{itemize}

\begin{figure*}[h!]
    \centering
    \hfill
    \begin{subfigure}[c]{0.5\textwidth}
    \centering
    \includegraphics[width=0.6\textwidth]{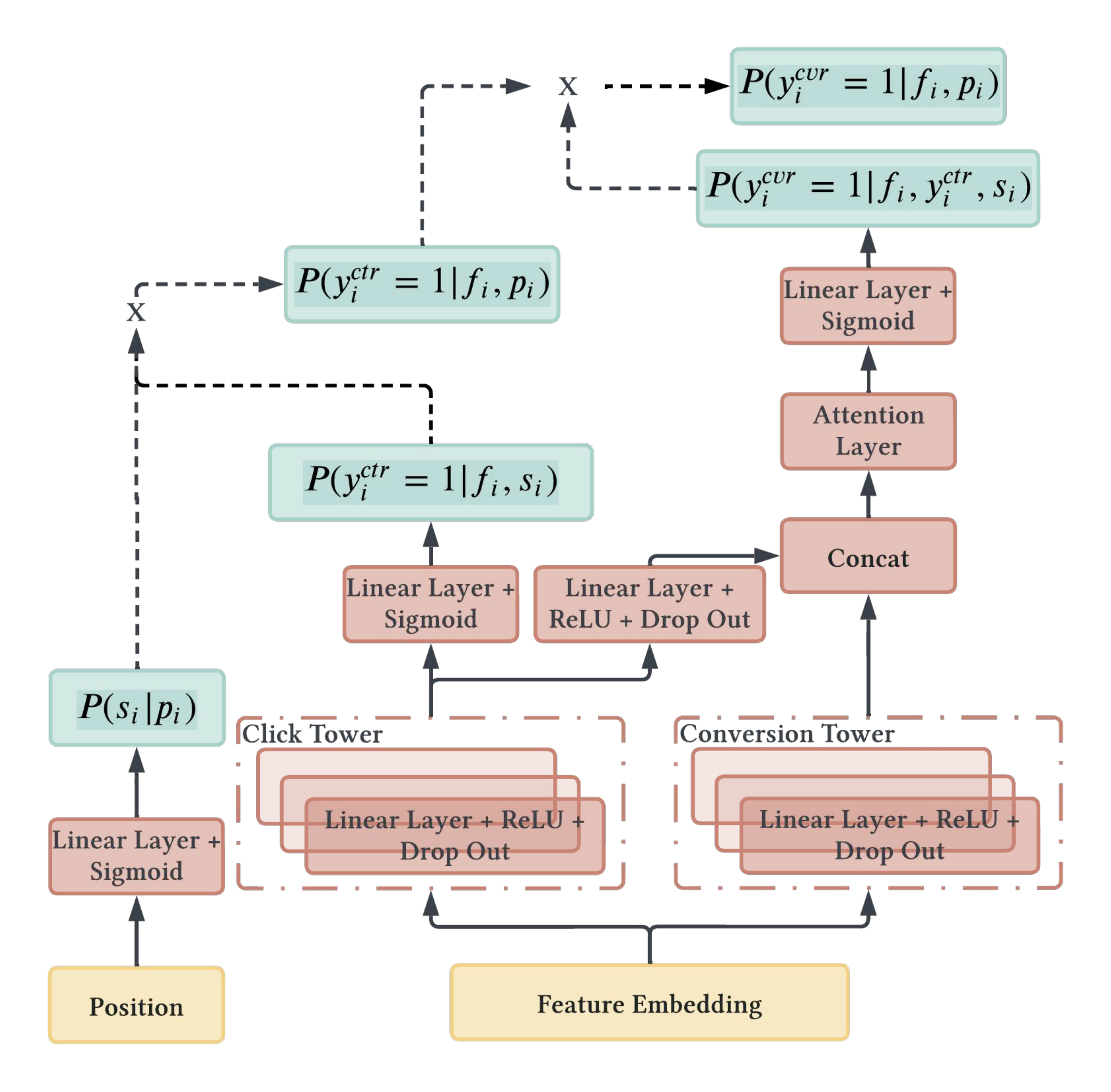}
    \caption{Position Aware Click-Conversion Model}
    \label{fig:model1}
    \end{subfigure}%
    \hfill
    \begin{subfigure}[c]{0.5\textwidth}
    \centering
    \includegraphics[width=0.6\textwidth]{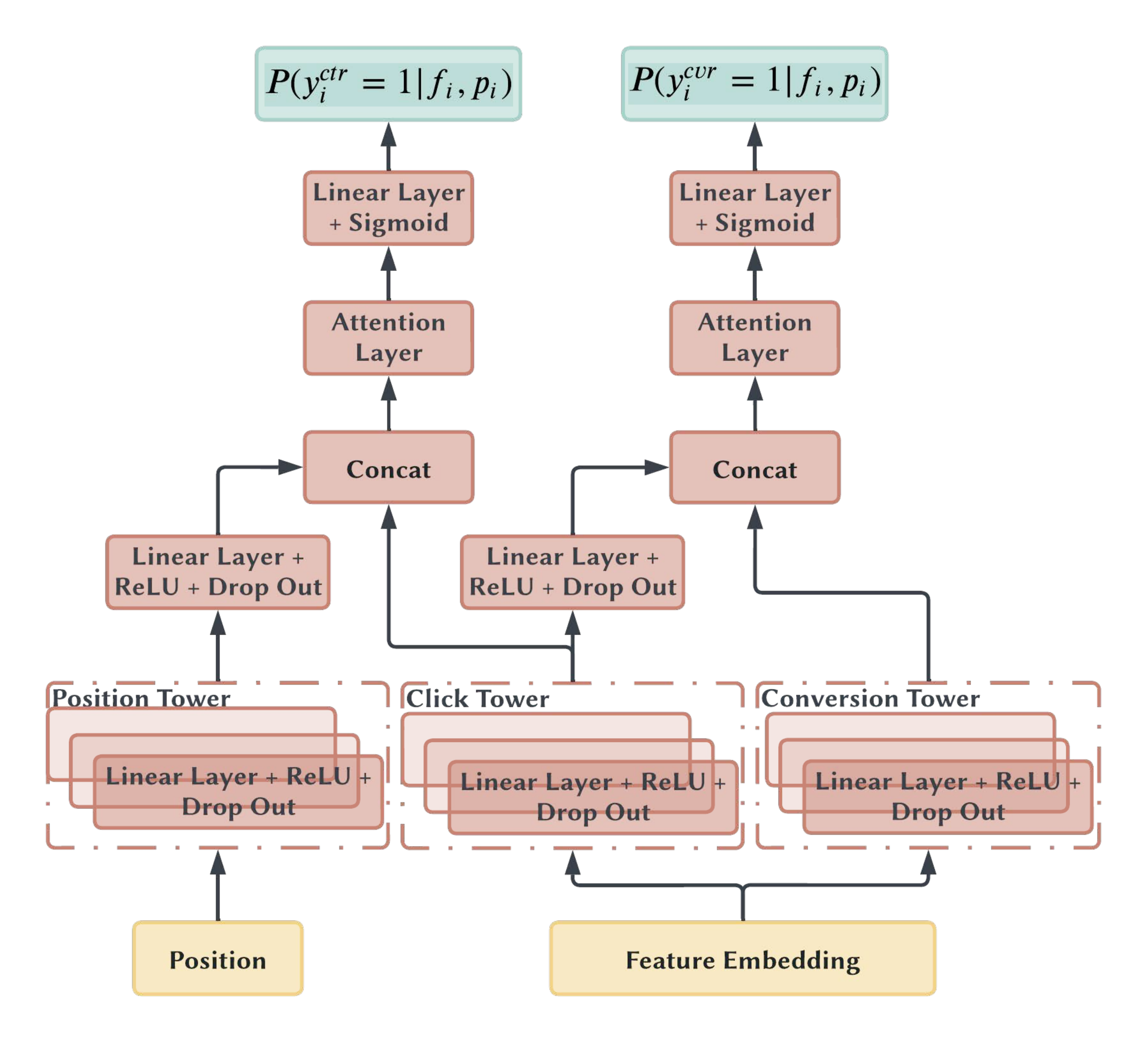}
    \caption{Position Aware Click-Conversion Model with Position Embedding}
    \label{fig:model2}
    \vspace{-0.46cm}
    \end{subfigure}%
    \vspace{-0.3cm}
    \caption{An overview of PACC and PACC-PE. 
    The left figure shows the overall structure of the PACC model which utilizes the assumptions and probability decomposition in \S\ref{ssect:theory} to estimate the probability of an item being clicked and the probability of an item being purchased based on position and feature embedding. The right figure illustrates the PACC-PE model that integrates neural networks to model product-specific position information as embedding.}
    \label{fig:models}
    \vspace{-0.3cm}
\end{figure*}

\vspace{-0.3cm}
\section{Methodology}
\label{sec: methods}
In this section, we first introduce our problem formulation and then present assumptions and theoretical basis. Finally, detailed explanations of our proposed PACC and PACC-PE models are provided.

\vspace{-0.1cm}
\subsection{Notation and Problem Formulation}
We assume the training set to be $T = \{(f_i, p_i) \rightarrow (y^{ctr}_i, y^{cvr}_i)\}|_{i=1}^N$, where
$f_i$ is features other than position of sample $i$, $p_i$ is the position of sample $i$, $y^{ctr}_i\in\{0,1\}$ is the click label of sample $i$,
$y^{cvr}_i\in\{0,1\}$ is the conversion label of sample $i$, 
and $N$ is the number of samples in $T$. 
Therefore, our models estimate the two probabilities: $P(y^{ctr}_i=1|f_i, p_i)$, the probability of an item to be clicked according to features and position, and $P(y^{cvr}_i=1|f_i, p_i)$, the probability of an item to be purchased according to features and position. Besides, we use $s_i\in\{0,1\}$ to represent whether an item is seen by the user or not.

\vspace{-0.1cm}
\subsection{Assumptions and Theoretical Basis}
\label{ssect:theory}
PACC is based on two assumptions: 
1) whether an item will be seen by the user is only related to item position, which is $P(s_i|f_i, p_i) = P(s_i|p_i)$; 
2) if an item is already seen, whether it will be clicked/purchased is independent of item position, which is $P(y^{ctr}_i|f_i, p_i, s_i=1) = P(y^{ctr}_i|f_i, s_i=1)$ and $P(y^{cvr}_i|f_i, p_i, s_i=1) = P(y^{cvr}_i|f_i, s_i=1)$. 

In addition to these two assumptions, two facts underlie PACC: 
1) an item has to be seen before it can be clicked, which is $P(y^{ctr}_i=1|f_i, p_i) = P(y^{ctr}_i=1 \cap s_i=1|f_i, p_i)$; 
2) an item has to be clicked and seen before it can be purchased, which is $P(y^{cvr}_i=1|f_i, p_i) = P(y^{cvr}_i=1 \cap y^{ctr}_i=1|f_i, p_i)$ and $P(y^{cvr}_i=1|f_i, p_i, y^{ctr}_i) = P(y^{cvr}_i=1 \cap s_i=1|f_i, p_i, y^{ctr}_i)$. 

The probability of an item being clicked
is represented as:
\begin{equation}
    \begin{aligned}
        P(y^{ctr}_i=1|f_i, p_i) 
        &= P(y^{ctr}_i=1|f_i, p_i, s_i=1) \cdot P(s_i=1|f_i, p_i)\\
        &= P(y^{ctr}_i=1|f_i, s_i=1) \cdot P(s_i=1|p_i).\\
    \end{aligned}
\label{eq:1}
\end{equation}

The probability of an item to be purchased is represented as:
\begin{equation}
    \begin{aligned}
        P(y^{cvr}_i=1|f_i, p_i)
        &= P(y^{cvr}_i=1 \cap y^{ctr}_i=1|f_i, p_i)\\
        &= P(y^{cvr}_i=1|f_i, y^{ctr}_i, s_i) \cdot P(y^{ctr}_i = 1|f_i, p_i).\\
    \end{aligned}
\label{eq:2}
\end{equation}

\vspace{-0.1cm}
\subsection{Proposed Framework}

\vspace{-0.05cm}
\subsubsection{Overview}
Figure~\ref{fig:models} illustrates the architecture of our proposed models. 
In both models, CTR prediction and CVR prediction share the same feature embedding but have different model architectures and parameters. A neural network is utilized to adaptively learn what and how much information to transfer from CTR prediction to CVR prediction. Position information is modeled separately and combined in innovative ways with the click-conversion multi-task model. With this novel framework, our proposed models can jointly alleviate the position bias of both CTR and CVR prediction.
\vspace{-0.05cm}
\subsubsection{Position Aware Click-Conversion Model}
The model architecture of PACC is shown in Fig. \ref{fig:model1}. 
Given the input feature $f_i$ embedded as $v_i$,
for task $k \in \{ctr, cvr \}$, the output of $k$ Tower is defined as $T_k = g^k_t(v_i)$,
where $g^k_t(\cdot)$ is three linear layers each followed by a ReLU activation function and a drop-out layer.
Then $T_{ctr}$ is fed into a linear layer with a sigmoid function to calculate $P(y^{ctr}_i|f_i, s_i=1)$, which is the probability of an item to be clicked after it is seen. $T_{ctr}$ is also fed into a linear layer followed by a ReLU activation function and a drop-out layer, whose output is $INFO_{ctr}$. Then $T_{cvr}$ and $INFO_{ctr}$ are concatenated and fed into an attention layer as $A_{cvr} = g^{cvr}_{a}([T_{cvr};INFO_{ctr}])$,
where $g^{cvr}_{a}$ is the function of the attention layer and $A_{cvr}$ is the output. Then $A_{cvr}$ is fed into a linear layer with a sigmoid function to calculate $P(y^{cvr}_i|f_i, y^{ctr}_i=1, s_i=1)$, which is the probability of an item to be purchased if it is already seen and clicked.

$P(s_i|p_i)$, the probability of an item to be seen given the position, is modeled using a linear layer and a sigmoid function. Then $P(s_i|p_i)$ is multiplied by $P(y^{ctr}_i|f_i, s_i=1)$ for $P(y_i^{ctr}=1|f_i, p_i)$ and $P(y_i^{ctr}=1|f_i, p_i)$ is multiplied by $P(y^{cvr}_i|f_i, y^{ctr}_i=1, s_i=1)$ for $P(y_i^{cvr}|f_i, p_i)$ according to Eq.(\ref{eq:1}) and Eq.(\ref{eq:2}).
\vspace{-0.05cm}
\subsubsection{Position Aware Click-Conversion Model with Position Embedding}
The architecture of PACC-PE is shown in Fig. \ref{fig:model2}. Given $v_i$ as the shared feature embedding of sample $i$, for task $k \in \{pos, ctr, cvr \}$, the output of $k$ Tower is defined as $T_k = g^k_t(v_i)$, 
where $g^k_t(\cdot)$ encodes $v_i$ through three linear layers each followed by a ReLU activation function and a drop-out layer.
$T_{pos}$ is then fed into a linear layer followed by a ReLU activation function and a drop-out layer, whose output is $INFO_{pos}$. Then $T_{ctr}$ and $INFO_{pos}$ are concatenated and fed into an attention layer as $A_{ctr} = g^{ctr}_{a}([T_{ctr};INFO_{pos}])$,
where $g^{ctr}_{a}$ is the function of the attention layer and $A_{ctr}$ is the output. Then $A_{ctr}$ is fed into a linear layer with a sigmoid function to calculate $P(y^{ctr}_i|f_i, p_i)$, which is the probability of an item to be clicked.
$P(y^{cvr}_i|f_i, p_i)$ is obtained similar as $P(y^{ctr}_i|f_i, p_i)$. 

The difference between PACC and PACC-PE is that PACC models position information into a scalar while PACC-PE models product-specific position information into an embedding. Thus, PACC-PE focuses on representing product-related position information, which is richer and more useful. Besides, PACC-PE has high fault tolerance compared with PACC, where subsequent tasks will be greatly affected if the probability of a former task is predicted wrongly.
\vspace{-0.05cm}
\subsubsection{Loss Function}
The loss function of our proposed models is
\begin{equation}
    \mathcal{L_(\theta)} = \mathcal{L}_{CTR}(\theta) + \mathcal{L}_{CVR}(\theta) + \mathcal{L}_{res}(\theta),
\label{eq:3}
\end{equation}
where $\mathcal{L}_{CTR}(\theta)$ is the binary cross entropy loss of the CTR prediction task, $\mathcal{L}_{CVR}(\theta)$ is the binary cross entropy loss of the CVR prediction task and $\mathcal{L}_{res}(\theta)$ is a restriction loss. The restriction loss is based on the fact that an item has to be clicked before it can be purchased, which is defined as
\begin{equation}
    \mathcal{L}_{res}(\theta) = \sum_{i=0}^{N} max(P(y_i^{ctr}|f_i, p_i) - P(y_i^{cvr}|f_i, p_i), 0).
\label{eq:4}
\end{equation}

\vspace{-0.3cm}
\section{Experiments}

\vspace{-0.1cm}
\subsection{Dataset and Data Preprocessing}
The training and testing query-item pair data are collected using Walmart\footnote{https://www.walmart.com} sponsored ads logs. The dataset is a real-world dataset with position bias, containing 4.2M training samples, 1.1M validation samples, and 7.5M testing samples. 

The extracted features are categorized into three types: categorical features, numeric features, and text features. 
The categorical features are transformed into one-hot vectors; the numeric features are normalized; and the text features are embedded using BERT~\cite{kenton2019bert} to calculate cosine similarity scores and element-wise product between query and ad item. 
The element-wise product obtained using $BERT_{base}$ has a dimension of 768, whereas the dimensions of other features are much smaller. To prevent the element-wise product from overwhelming other features, we use PCA to reduce its dimensionality to 5.
In addition, the position feature is transformed into a one-hot format.


\vspace{-0.1cm}
\subsection{Evaluation Metrics}
Since the standard Mean Reciprocal Rank (MRR) assumes that the clicked/purchased items are relevant while ignoring the position bias, we use weighted MRR \cite{metrics} to evaluate ranking effectiveness, which is formulated as $MRR^{m} = \frac{1}{\sum_{i=1, \tilde{y}_i^m=1}^N w_i^m} \sum_{i=1, \tilde{y}_i^m=1}^N w_i^m \frac{1}{p_i}$,
where $m\in \{ctr, cvr\}$, $N$ is the number of samples, $w_i^m = \frac{1}{P(s_i|f_i, p_i)}$ is a weight of $i$-th sample, $p_i$ is the position of the $i$-th sample and $\tilde{y}_i^m$ is the predicted label. 
$w_i^m$ is defined through the reciprocal of $P(s_i|p_i)$ which reflects the impact of position.

For PACC, $w_i^m$ can be calculated in a straightforward way since $P(s_i|p_i)=P(s_i|f_i, p_i)$ is obtained after training.

To compute $P(s_i|f_i, p_i)$ for PACC-PE as mentioned in \cite{ubiased_ltr}, we first swap $p_i$ with $p_i=r$ for item $i$. 
The probability of this item being clicked at position $p_i=r$ is
\begin{equation}
    \label{eq:9}
    P(y_i^{ctr}=1|f_i, p_i=r) = P(y_i^{ctr}=1|f_i, s_i=1)\cdot P(s_i|f_i, p_i=r).
\end{equation}
The probability of this item being clicked at position $p_i$ is
\begin{equation}
    \label{eq:10}
    P(y_i^{ctr}=1|f_i, p_i) = P(y_i^{ctr}=1|f_i, s_i=1) \cdot P(s_i|f_i, p_i).
\end{equation}
With Eq. (\ref{eq:9}) and Eq. (\ref{eq:10}), the ratio $\frac{P(s_i|f_i, p_i)}{P(s_i|f_i, p_i=r)} = \frac{P(y_i^{ctr}=1|f_i, p_i)}{P(y_i^{ctr}=1|f_i, p_i=r)}$ is obtained.
Thus, the weight $w_i^{ctr}$ for PACC-PE is
\begin{equation}
    \label{eq:w_pacc-pe}
    w_i^{ctr} = \frac{P(y_i^{ctr}=1|f_i, p_i=r)}{P(y_i^{ctr}=1|f_i, p_i)} \cdot \frac{1}{P(s_i|f_i, p_i=r)},
\end{equation}
which is proportional to $\frac{P(y_i^{ctr}=1|f_i, p_i=r)}{P(y_i^{ctr}=1|f_i, p_i)}$ since $P(s_i|p_i=r)$ is a constant for a fixed position $r$. $w_i^{cvr}$ can be calculated similarly.

In addition to the weighted MRR, we also apply the widely-used evaluation metrics: standard MRR, AUC, and position-wise AUC (PAUC) \cite{huang2021metrics2} to evaluate model performance.



\vspace{-0.1cm}
\subsection{Experimental Results and Analysis}

\begin{table}[t!]
\caption{Ranking Performance Comparison Results with Based Models}
\vspace{-0.3cm}
\centering
\resizebox{\linewidth}{!}{
\begin{tabular}{l|c|c|c|c|c|c|c|c}
\hline 
\multirow{2}*{\textbf{Models}} & \multicolumn{4}{c|}{CTR} & \multicolumn{4}{c}{CVR} \\
\cline{2-9}
~ & \textbf{Weighted MRR} & \textbf{MRR} & \textbf{PAUC} & \textbf{AUC} & \textbf{Weighted MRR} & \textbf{MRR} & \textbf{PAUC} & \textbf{AUC} \\
\hline
DMT~\cite{dmt} & 39.65 & \textbf{39.68} & 88.43 & 87.75 & 42.38 & 42.31 & 56.60 & 83.82 \\
PAL~\cite{guo2019pal} & \textbf{39.67} & 39.44 & 88.67 & 88.02 & 42.52 & 43.37 & 55.22 & 80.82 \\
AITM~\cite{AITM} & 39.40 & 39.25 & 88.89 & 89.04 & 40.64 & 43.18 & 60.39 & 89.56 \\
PACC & 39.56 & 39.35 & 88.96 & 88.28 & 43.73 & 43.53 & \textbf{60.79} & \textbf{90.05}  \\
PACC-PE & 39.43 & 39.43 & \textbf{92.16} & \textbf{91.68} & \textbf{47.44} & \textbf{47.44} & 60.72 & 89.28 \\
\hline
\end{tabular}
}
\label{tab:experimental results}
\vspace{-0.6cm}
\end{table}

\subsubsection{Ranking Effectiveness}
We compare the weighted MRR, MRR, PAUC, and AUC of our proposed models and the baselines PAL~\cite{guo2019pal}, AITM~\cite{AITM}, and DMT~\cite{dmt} models. 
PAL is a single-task position bias mitigation model; AITM is a multi-task model; DMT is a multi-task position bias mitigation model.
The evaluation results are reported in Table~\ref{tab:experimental results}. 

For CTR prediction, all models perform similarly in terms of weighted MRR and MRR. PACC-PE outperforms the best baseline by 3.27\%/2.64\% in terms of PAUC/AUC with p-value = 0.0072 and confidence level > 99\%. 
For CVR prediction, PACC-PE significantly outperforms all other baseline models regarding weighted MRR and MRR, increasing weighted MRR/MRR by 4.92\%/4.07\% with p-value = 0.0088 and confidence level > 99\%. PACC also outperforms other baseline models. 


The substantial improvement in CVR prediction of our proposed models indicates that jointly mitigating position bias for both CTR and CVR and considering the sequential dependencies in order between click and purchase is effective for performance improvement.

\subsubsection{Position Bias}
\begin{figure}[b!]
\vspace{-0.5cm}
    \centering
    \begin{subfigure}[b]{0.158\textwidth}
    \centering
    \includegraphics[width=\textwidth]{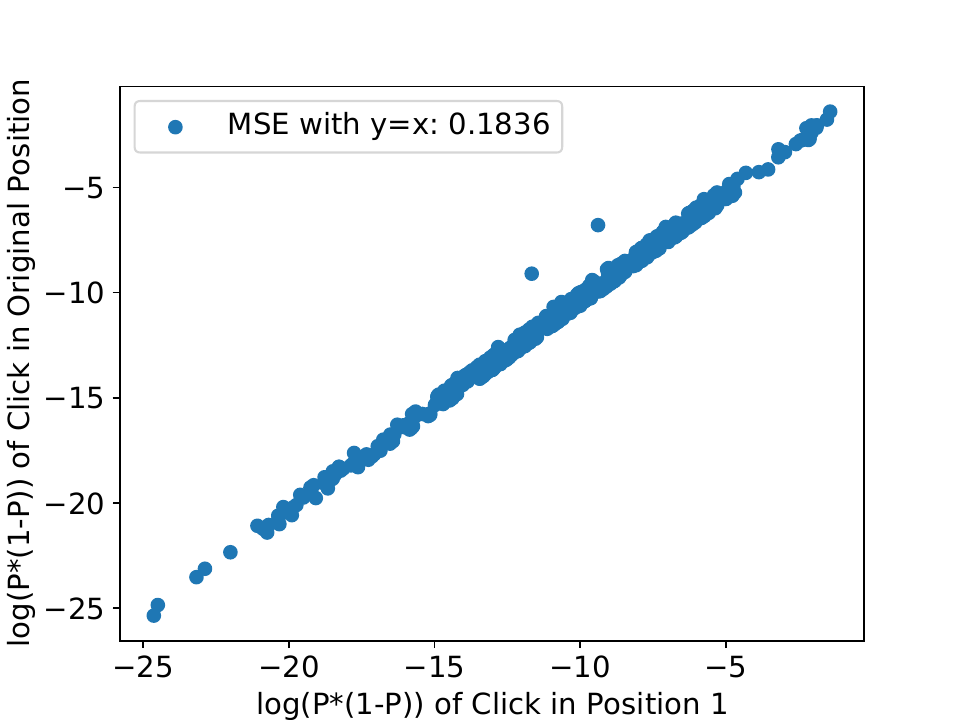}
    \caption{AITM}
    \label{fig:AITM_ctr_log}
    \end{subfigure}%
    \hfill
    \begin{subfigure}[b]{0.158\textwidth}
    \centering
    \includegraphics[width=\textwidth]{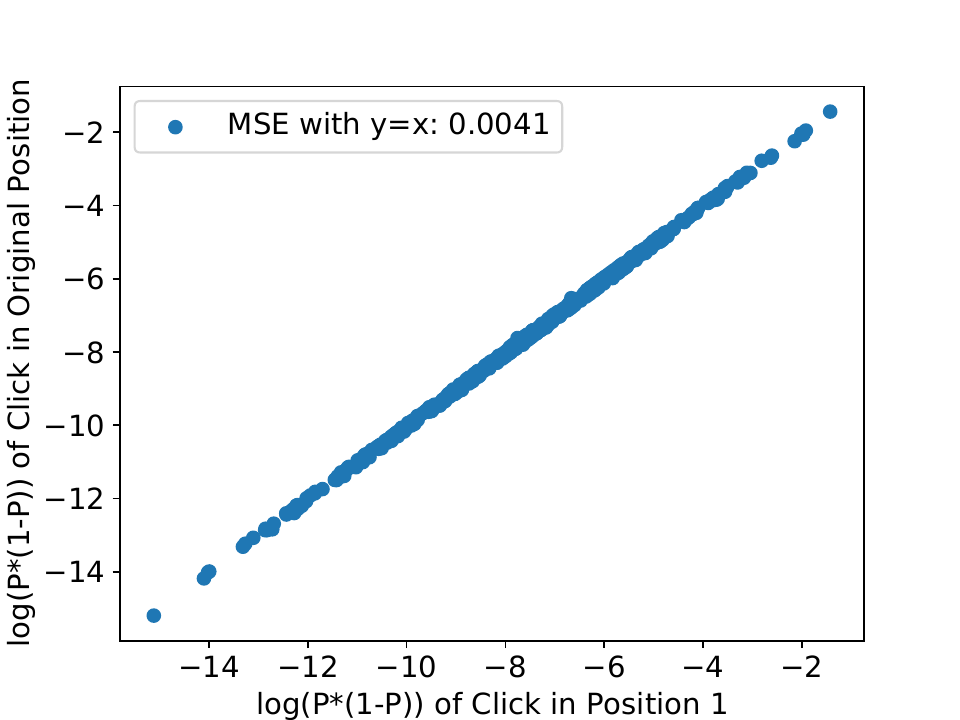}
    \caption{PACC}
    \label{fig:AITM_prob_ctr_log}
    \end{subfigure}%
    \hfill
    \begin{subfigure}[b]{0.158\textwidth}
    \centering
    \includegraphics[width=\textwidth]{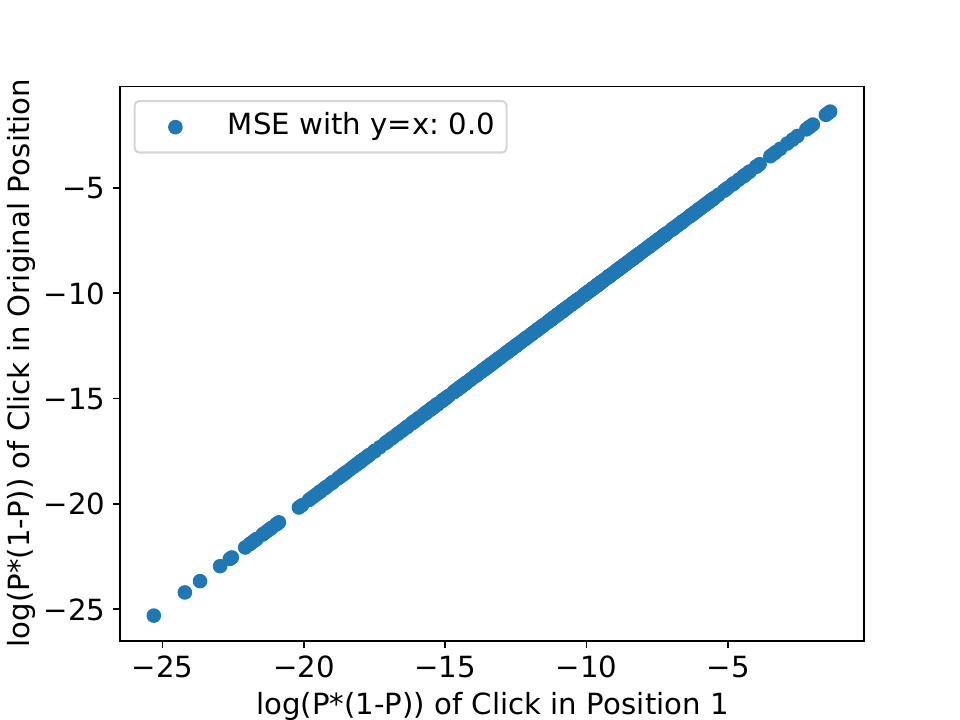}
    \caption{PACC-PE}
    \label{fig:AITM_ait_ctr_log}
    \end{subfigure}%
    \caption{The log odds of probabilities of being \textbf{clicked} of 500 randomly selected query-item pairs.
    The y-axis is the probability of being clicked at the original position; the x-axis is the probability of being clicked at position \textbf{1}.}
    \label{fig:three_models}
    \vspace{-0.5cm}
\end{figure}

\begin{figure}
    \centering
    \begin{subfigure}[b]{0.158\textwidth}
    \centering
    \includegraphics[width=\textwidth]{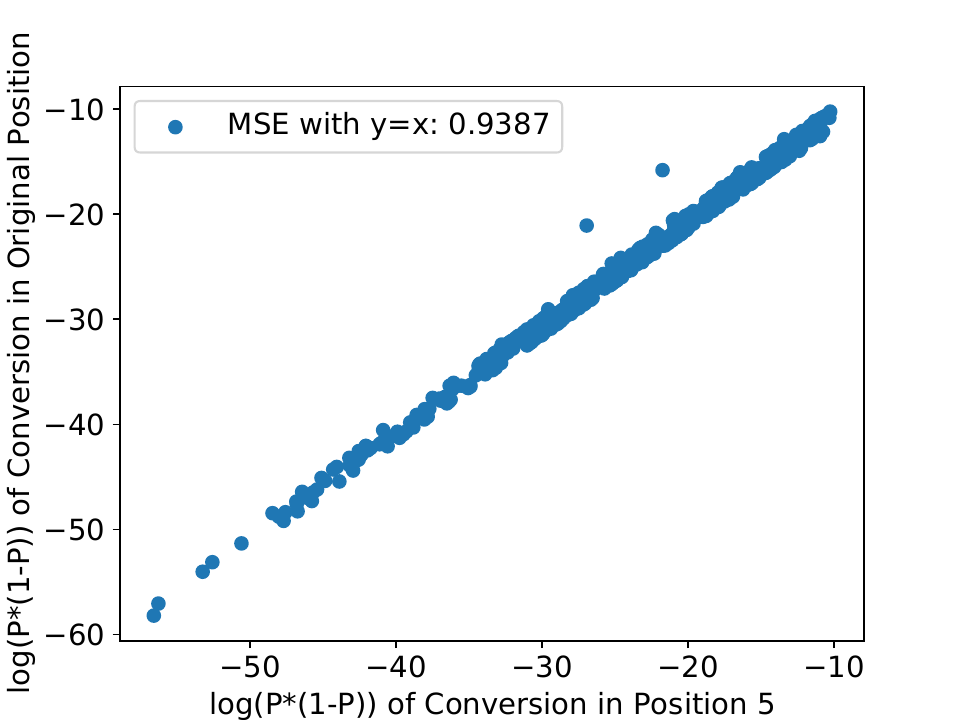}
    \caption{AITM}
    \label{fig:AITM_cvr_log}
    \end{subfigure}%
    \hfill
    \begin{subfigure}[b]{0.158\textwidth}
    \centering
    \includegraphics[width=\textwidth]{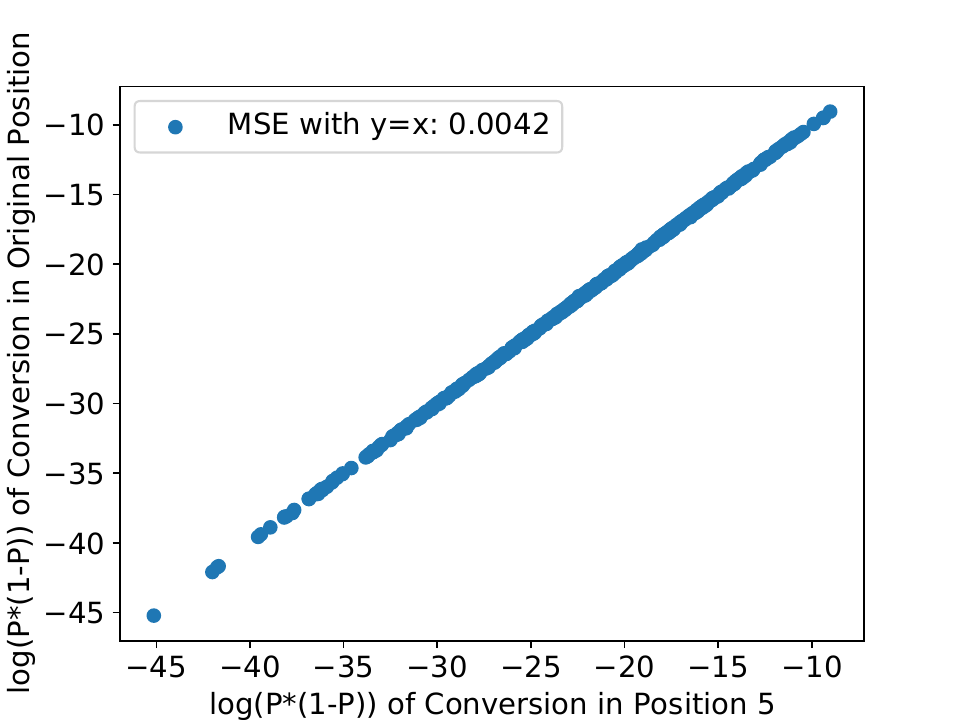}
    \caption{PACC}
    \label{fig:AITM_prob_cvr_log}
    \end{subfigure}%
    \hfill
    \begin{subfigure}[b]{0.158\textwidth}
    \centering
    \includegraphics[width=\textwidth]{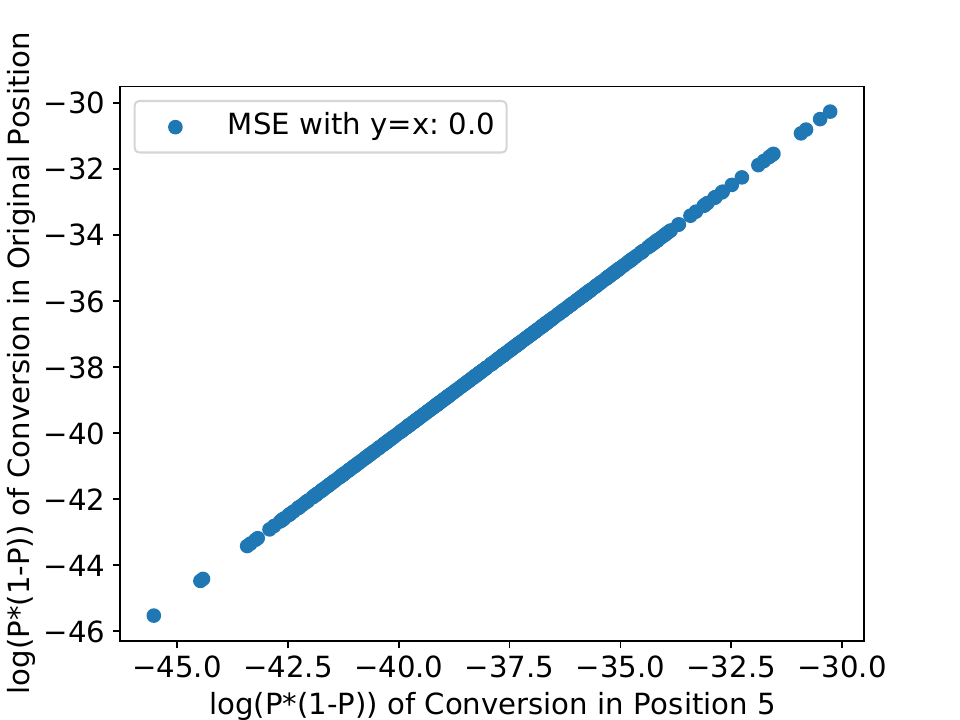}
    \caption{PACC-PE}
    \label{fig:AITM_ait_cvr_log}
    \end{subfigure}%
    \vspace{-0.3cm}
    \caption{The log odds of probabilities of being \textbf{purchased} of 500 randomly selected query-item pairs.
    The y-axis is the probability of being purchased at the original position; the x-axis is the probability of being purchased at position \textbf{1}.}
    \label{fig:three_models2}
    \vspace{-0.5cm}
\end{figure}

To generally evaluate the ability of our proposed model in mitigating position bias, we randomly select 500 query-item pairs and visualize their probability of being clicked and purchased before and after swapping positions. To better interpret the predicted probabilities, the log odds function is used to project probabilities in log odds space. From Fig~\ref{fig:three_models} and Fig~\ref{fig:three_models2}, the sample points of PACC and PACC-PE fit better on $y=x$ compared to AITM, indicating that swapping item positions with position 1 has less influence on the predicted probabilities of being clicked and purchased by PACC and PACC-PE.

\begin{figure}
\vspace{-0.1cm}
     \centering
     \begin{subfigure}[b]{0.158\textwidth}
         \centering
         \includegraphics[width=\textwidth]{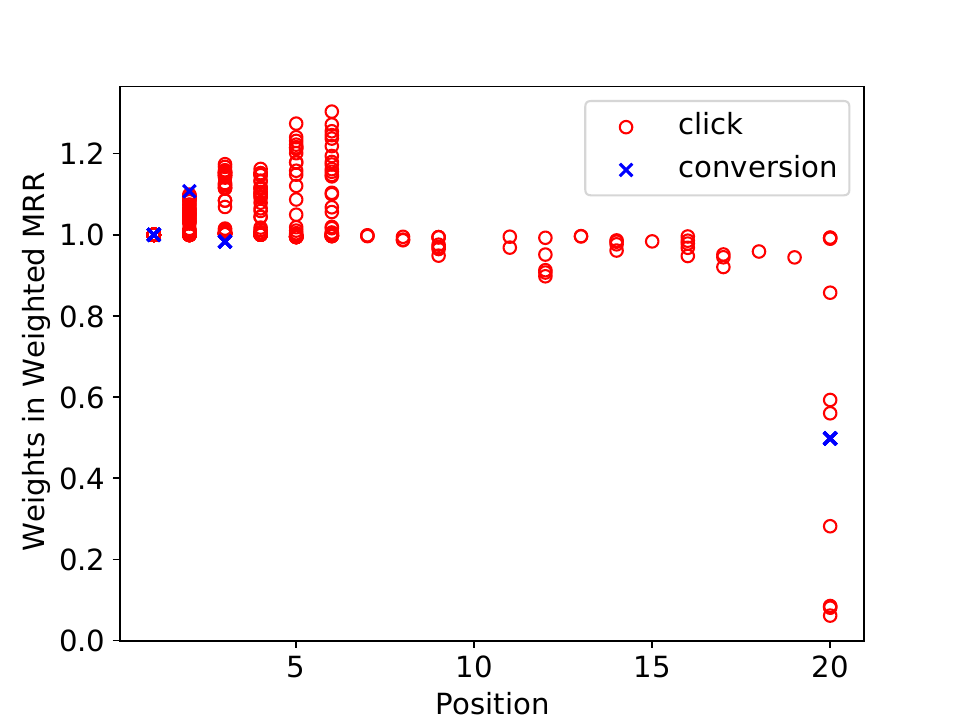}
         \caption{AITM}
         \label{fig:AITM}
     \end{subfigure}%
     \hfill
     \begin{subfigure}[b]{0.158\textwidth}
         \centering
         \includegraphics[width=\textwidth]{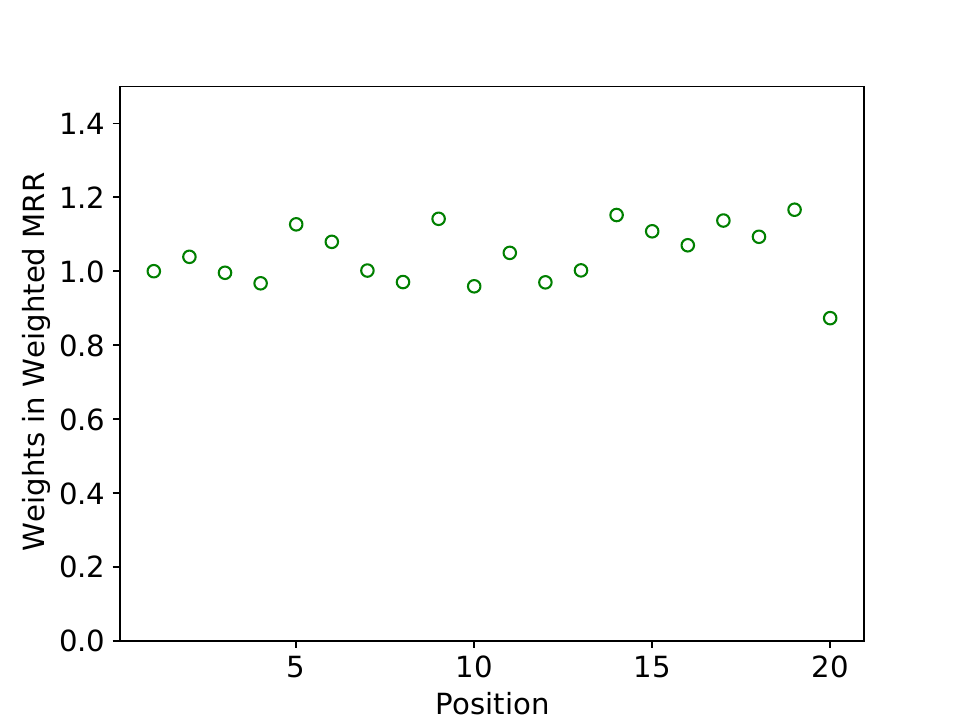}
         \caption{PACC}
         \label{fig:PAPTM}
     \end{subfigure}%
     \hfill
     \begin{subfigure}[b]{0.158\textwidth}
         \centering
         \includegraphics[width=\textwidth]{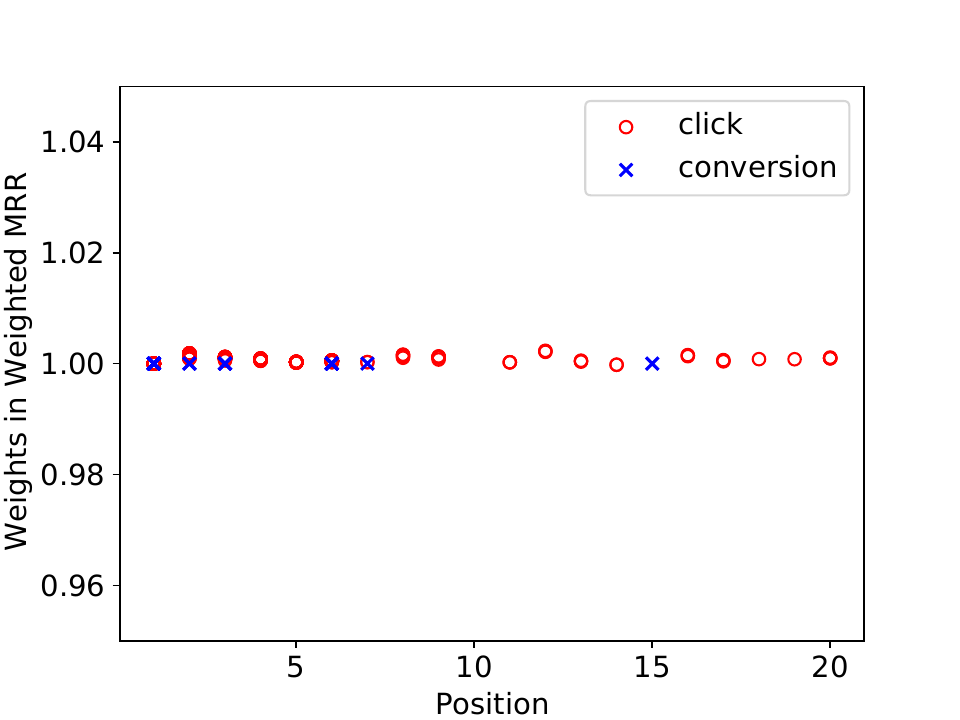}
         \caption{PACC-PE}
         \label{fig:PAATM}
     \end{subfigure}%
    \vspace{-0.3cm}
    \caption{Visualization of the impact of position swapping demonstrated by 300 randomly selected examples. Red $\circ$ are clicked query-item pairs. Blue $\times$ are purchased query-item pairs. Green $\circ$ are the impact of swapping positions at each position in PACC.
    }
    \label{fig:AITM_AITM_ait}
    \vspace{-0.5cm}
\end{figure}

\subsubsection{Position Bias on Different Positions}
To evaluate the ability of our proposed models in mitigating position bias on different positions, we investigate the differences in model prediction changes for items at different positions due to swapping positions with position 1. The smaller the position bias, the smaller the prediction changes due to swapping positions.
The impact of swapping item positions is visualized in Fig~\ref{fig:AITM_AITM_ait}. 
In Fig~\ref{fig:AITM} and Fig~\ref{fig:PAATM}, $\frac{P(y_i^{ctr}=1|f_i, p_i=1)}{P(y_i^{ctr}=1|f_i, p_i)}$ and $\frac{P(y_i^{cvr}=1|f_i,p_i=1)}{P(y_i^{cvr}=1|f_i, p_i)}$ are used to measure the impact of swapping item positions on CTR prediction and CVR prediction, respectively. As in Fig~\ref{fig:PAATM}, swapping item positions almost has no effect on both CTR and CVR prediction for PACC-PE on all positions, indicating that PACC-PE has the ability to mitigate position bias on all positions. In Fig~\ref{fig:AITM} swapping item positions has a large impact on the top few items and the last item in the ranking list for AITM. For items on the top, swapping them to position 1 improves the probability of being clicked and purchased, while for the last items swapping them to position 1 decreases the probability of being clicked and purchased. This phenomenon is counter-intuitive because intuitively swapping items to position 1 should increase the probability of being clicked and purchased. 
One possible explanation is that 
some items are safe choices, but not preferred. For these items, if they are ranked high, users will not click on them. But if they are ranked low, users may click them given no better items at the end of their browsing.
In Fig~\ref{fig:PAPTM}, $\frac{P(s_i| p_i=1)}{P(s_i| p_i)}$ is used to measure the impact of swapping item positions. The impact of swapping item positions for PACC is low compared to AITM, demonstrating the ability of PACC to alleviate position bias. For items at most positions, swapping positions to position 1 increases the probability of being clicked and purchased a little bit.

One difference between PACC-PE and PACC from Fig~\ref{fig:PAATM} and Fig~\ref{fig:PAPTM} is that for PACC, position bias of different items at the same position is the same, while for PACC-PE position bias of different items at the same position is different. This difference makes PACC-PE more flexible to different items and conveys richer information.

\vspace{-0.3cm}
\section{Conclusion}
To jointly mitigate position bias that exists in both item CTR and CVR prediction, we propose two position-bias-free CTR and CVR prediction models: Position Aware Click-Conversion and PACC with Position Embedding. In PACC, the position is modeled as a probability while in PACC-PE position is modeled into embedding. Our experiments and analyses illustrate that our proposed models achieve better ranking effectiveness than the state-of-the-art models and effectively mitigate position bias in all positions. Besides, PACC-PE outperforms PACC in ranking effectiveness and position debias due to the rich information by modeling product-specific position information as embedding.



\bibliographystyle{ACM-Reference-Format}
\balance
\bibliography{sigir_short}

\appendix



\end{document}